\documentclass[a4paper,11pt,dvips]{article}
\usepackage{graphicx}
\usepackage{amsmath}
\usepackage{amssymb}
\usepackage{latexsym}
\usepackage[colorlinks]{hyperref}
\usepackage{color}
\usepackage{float}
\usepackage{cite}
\usepackage{makeidx}
\usepackage{colortbl}

\newcommand{\bra}{\begin{array}}
\newcommand{\era}{\end{array}}
\newcommand{\beq}{\begin{equation}}
\newcommand{\eeq}{\end{equation}}
\newcommand{\bqr}{\begin{eqnarray}}
\newcommand{\eqr}{\end{eqnarray}}

\def\BC{\bb C}
\def\_\BC{\bbi C}

\def \lp {\left(}
\def \gp {\right)}

\def\no2 {{\textstyle{n\over 2}}}

\newcommand{\lb}{\label}

\begin{document}
\begin{titlepage}
\setcounter{page}{1}
\renewcommand{\thefootnote}{\fnsymbol{footnote}}

\begin{flushright}
\end{flushright}

\vspace{5mm}
\begin{center}

{\Large \bf {Transmission via Triangular Double Barrier
 \\ and Magnetic Fields in Graphene }}

\vspace{5mm}
{\bf Miloud Mekkaoui}$^{a}$,
 {\bf Ahmed Jellal\footnote{\sf ajellal@ictp.it --
a.jellal@ucd.ac.ma}}$^{a,b}$ and {\bf Hocine Bahlouli}$^{b,c}$

\vspace{5mm}

{$^{a}$\em Theoretical Physics Group,  
Faculty of Sciences, Choua\"ib Doukkali University},\\
{\em PO Box 20, 24000 El Jadida, Morocco}

{$^b$\em Saudi Center for Theoretical Physics, Dhahran, Saudi Arabia}

{$^c$\em Physics Department,  King Fahd University
of Petroleum $\&$ Minerals,\\
Dhahran 31261, Saudi Arabia}

\vspace{3cm}

\begin{abstract}
We study the transmission probability of Dirac fermions in graphene scattered by
a triangular double barrier potential in the presence of an external magnetic field.
Our system made of two triangular potential barrier regions separated by a well region
characterized by an energy gap $G_p$. Solving our Dirac-like equation and matching the solutions
at the boundaries we express our transmission and reflection coefficients in terms of transfer matrix.
We show in particular that the transmission exhibits oscillation resonances that are manifestation of
the Klein tunneling effect.

\end{abstract}
\end{center}

\vspace{3cm}


\end{titlepage}


\section{Introduction}

Graphene~\cite{Novoselov} remains among the most fascinating and
attractive subject has been seen right now in condensed matter physics. This is
because of its exotic physical properties and the apparent
similarity of its mathematical model to the one describing
relativistic fermions in two dimensions. As a consequence of this
relativistic-like behavior particles could tunnel through very high
barriers in contrast to the conventional tunneling of
non-relativistic particles, an effect known in relativistic field
theory as Klein tunneling. This tunneling effect has already been
observed experimentally~\cite{Stander} in graphene systems. There
are various ways for creating barrier structures in
graphene~\cite{Katsnelsonn, Sevinçli}. For instance, it can be done
by applying a gate voltage, cutting the graphene sheet into finite
width to create a nanoribbons, using doping or through the
creation of a magnetic barrier. In the case of graphene, results
of the transmission coefficient and the tunneling conductance were
already reported for the electrostatic barriers~\cite{Sevin,
Masir, DellAnna, Mukhopadhyay}, magnetic barriers~\cite{DellAnna, Choubabi, Mekkaoui},
potential barrier~\cite{Jellal, Alhaidari} and triangular barrier~\cite{HBahlouli}.


We study the transmission probability of Dirac fermions in graphene scattered
by a triangular double barrier
potential in the presence of an inhomogeneous magnetic fields $B$.
We emphasis that  $B$-field discussed in
our manuscript is applied externally. It can be created for
instance by depositing a type-I superconducting film on top of the
system and remove a strip $|x|<d_1$ of the superconductor and
apply a perpendicular magnetic field. This patterning technique of
creating the desired magnetic field profile was proposed in
\cite{Matulis}. One of the interesting features of such
inhomogeneous magnetic field profile is that it can bind
electrons, contrary to the usual potential step. Such a step
magnetic field will indeed result in electron states that are
bound to the step $B$-field and that move in one direction along the
step. Thus there is a current along the $y$-direction but it is a
very small effect and is not relevant for our problem (those
electrons have $k_{x} = 0$). Indeed, we consider free electron
states that have in general $k_x$ non zero, because otherwise they
will not tunnel. A recent work studied double barriers with
magnetic field in graphene without mass term \cite{Ramezani}.



The paper is organized as follows. In section 2, we formulate our
model by setting the Hamiltonian system describing particles
scattered by a triangular double barrier  whose well potential zone is
subject to a magnetic field with a mass term. In section 3, we
consider the case of static double barriers and
derive the energy spectrum to finally  
determine the transmission and reflection probabilities. Their
behaviors are numerically investigated and in particular resonances were seen in different regions
as well as the Klein tunneling effect. In section 4, we study the
same system but this time by taking into account the presence of an
inhomogeneous magnetic field. 
Using boundary conditions, we  split the energy into three domains
and then calculate the transmission probability in each case.
In each situation, we discuss the transmission at resonances that
characterize each region and stress the importance of our
results. We conclude our work in the final section.

\section{ Mathematical model}

We consider a system of massless Dirac fermions incident on a two-dimensional strip of
graphene having energy $E$ and at incidence angle
$\phi_1$ with respect to the $x$-direction. This system
is a flat sheet of graphene subject to a square potential barrier
along the $x$-direction while particles are free in the
$y$-direction. Let us first describe the
geometry of our system, which is made of five regions denoted by
${\sf{j = 1}}, {\sf{\cdots}}, {{\sf5}}$. Each region is characterized by its constant
potential and interaction with external sources. All
regions are formally described by a Dirac-like Hamiltonian
\begin{equation}\lb{Ham1}
H=v_{F}
{\boldsymbol{\sigma}}\cdot\left(\textbf{p}+\frac{e}{c}\textbf{A}\right)+
V(x){\mathbb
I}_{2}+G_p\Theta\left(d_{1}^{2}-x^{2}\right)\sigma_{z}
\end{equation}
where {${v_{F}\approx 10^6 m/s}$  is the Fermi
velocity, ${{\boldsymbol{\sigma}}=(\sigma_{x},\sigma_{y})}$ and
$\sigma_{z}$} are the Pauli matrices in pseudospin space,
$\textbf{p}=-i\hbar(\partial_{x},
\partial_{y})$ is the momentum operator, ${\mathbb I}_{2}$ the $2 \times 2$ unit matrix,
 $V(x)=V_{\sf j}$ is the electrostatic potential in the ${\sf j}$-th scattering region
 and $\Theta$ is
the Heaviside step function.
The magnetic field $B(x, y)= B(x)$ is defined through the Landau gauge,
which allows the vector potential to be of the form $\textbf{A} =
(0,A_{y}(x))$ with $\partial_{x}A_{y}(x)= B(x)$. The parameter $G_p = m v_{F}^2$ is
the energy gap owing to the sublattice symmetry breaking,  it can also be
seen as the energy gap $G_p = G_{p, so}$ originating from
spin-orbit interaction.

First let us specify potential configuration that will constitute our double barrier potential
\begin{equation}\lb{popro}
V(x)=
\left\{%
\begin{array}{ll}
    \Lambda ( d_2 + \gamma x ) , & \hbox{$d_{1}\leq |x|\leq d_{2}$} \\
    V_{2}, & \hbox{$ |x|\leq d_{1}$} \\
    0, & \hbox{otherwise} \\
\end{array}%
\right.
\end{equation}
where {$\gamma=\pm1$,  $\gamma=1$ for $x\in [-d_2,
-d_1]$, $\gamma=-1$ for $x\in [d_1, d_2]$ and} the parameter
$\Lambda$ defined by $ \Lambda = \frac{V_1}{d_2-d_1}$ gives the
slope of triangular potentials. The graphical representation of
this potential is shown in Figure 1.
We define each potential region as follows: ${\sf{j = 1}}$ for $x \leq -d_2 $, ${\sf{j = 2}}$ for $ -d_2 \leq x \leq -d_1 $,
${\sf{j = 3}}$ for $ -d_1 \leq x \leq d_1 $, ${\sf{j = 4}}$ for $ d_1 \leq x \leq d_2 $ and ${\sf{j = 5}}$ for $ x \geq d_2 $. The corresponding
constant potentials are defined by \eqref{popro} and are denoted by $V_j$ in the $j$-th region.
\begin{figure}[ht]
  \centering
  \includegraphics[width=12cm, height=5cm ]{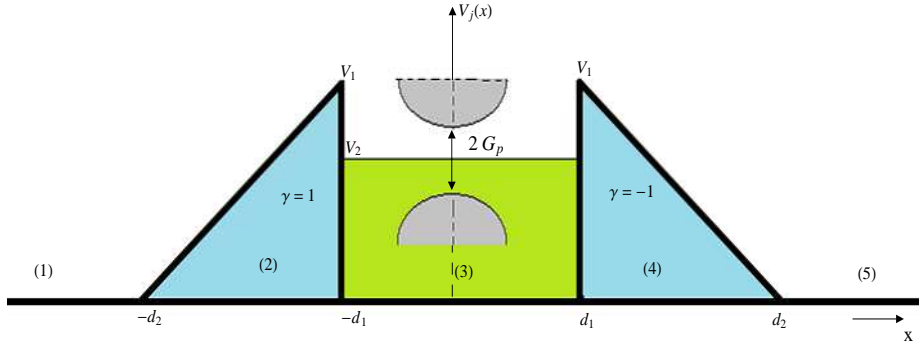}\\
  \caption{\sf Schematic diagram for the monolayer graphene double barrier.}\label{db.1}
\end{figure}

\section{Static double barrier}


We consider the Hamiltonian describing Dirac fermions in graphene scattered by an
electrostatic double barrier potential without magnetic field $\textbf{A} = 0$. In this case
\eqref{Ham1} reduces to
\begin{equation} \lb{eqh1}
H_{s}=v_{F} {\boldsymbol{\sigma}}\cdot\textbf{p}+V (x){\mathbb
I}_{2}+G_p\Theta\left(d_{1}^{2}-x^{2}\right)\sigma_{z}
\end{equation}
where ${\sf{j}}$ labels the five regions indicated schematically in
Figure \ref{db.1}  showing the space configuration of the
potential profile.  Due to sublattice symmetry we therefore need to study our system only near the \textbf{K}
point. The time-independent Dirac equation for the spinor
$\Phi(x,y)=\left(\varphi^{+},\varphi^{-}\right)^{T}$ at energy
$E=v_{F}\epsilon$ then reads, in the unit system $\hbar = m = c= 1$, as
\begin{equation} \lb{eqh1}
\left[{\boldsymbol{\sigma}}\cdot\textbf{p}+{v_j}{\mathbb
I}_{2}+\mu\Theta\left(d_{1}^{2}-x^{2}\right)\sigma_{z}\right]\Phi(x,y)=\epsilon
\Phi(x,y)
\end{equation}
where {$V_{\sf j}=v_{F}v_{\sf j}$ and $G_{p}=v_{F}\mu$}. Our
system is supposed to have finite width $W$ with infinite mass
boundary conditions on the wavefunction at the boundaries $y = 0$
and $y = W$ along the $y$-direction \cite{Tworzydlo, Berry}. These
boundary conditions result in a quantization of the transverse
momentum along the $y$-direction as
\begin{equation}
k_{y}=\frac{\pi}{W}\left(n+\frac{1}{2}\right),\qquad n=0,1,2 \cdots.
\end{equation}

One can therefore assume a spinor solution of the following form
$\Phi_{\sf j}=\left(\varphi_{\sf j}^{+}(x),\varphi_{\sf
j}^{-}(x)\right)^{\dagger}e^{ik_{y}y}$ and the subscript ${\sf
j}= 1, 2, 3, 4, 5$ indicates the space region while the
superscripts indicate the two spinor components. Solving the
eigenvalue equation to obtain the upper and lower components of
the eignespinor in the incident and reflection region {\sf 1} ($x
< - d_{2}$)
\begin{eqnarray}\label{eq3}
    && \Phi_{\sf 1}=  \left(
            \begin{array}{c}
              {1} \\
              {z_{1}} \\
            \end{array}
          \right) e^{i(k_{1}x+k_{y}y)} + r_{s,n}\left(
            \begin{array}{c}
              {1} \\
              {-z_{1}^{-1}} \\
            \end{array}
          \right) e^{i(-k_{1}x+k_{y}y)}\\
&& z_{1} =s_{1}\frac{k_{1}+ik_{y}}{\sqrt{k_{1}^{2}+k_{y}^{2}}}
\end{eqnarray}
where the sign function is defined by $s_{\sf j}={\mbox{sign}}{\left(E\right)}$.
The corresponding dispersion relation takes the form 
\beq
\epsilon=s_1\sqrt{k_1^2 +k_y^2}.
\eeq
In regions {\sf 2} and {\sf 4} ($d_{1}<|x|<d_{2}$),  the general
solution can be expressed in terms of the parabolic cylinder
function \cite{Abramowitz, Gonzalez, HBahlouli} as
\begin{equation}\lb{hiii1}
 \chi_{\gamma}^{+}=c_{n1}
 D_{\nu_n-1}\left(Q_{\gamma}\right)+c_{n2}
 D_{-\nu_n}\left(-Q^{*}_{\gamma}\right)
\end{equation}
where {$c_{n1}$ and $c_{n2}$ are constants,
$\nu_n=\frac{ik_{y}^{2}}{2\varrho}$ and $
Q_{\gamma}(x)=\sqrt{\frac{2}{\varrho}}e^{i\pi/4}\left(\gamma
\varrho x+\epsilon_{0}\right)$, with 
$\epsilon_{0}=\epsilon-v_{1}$, $\Lambda=v_{F}\varrho$,
$V_{1}=v_{F}v_{1}$}. The lower spinor component is given by
\begin{eqnarray}\lb{hiii2}
\chi_{\gamma}^{-}=-\frac{c_{n2}}{k_{y}}\left[
2(\epsilon_{0}+\gamma \varrho x)
 D_{-\nu_n}\left(-Q^{*}_{\gamma}\right)
+
 \sqrt{2\varrho}e^{i\pi/4}D_{-\nu_n+1}\left(-Q^{*}_{\gamma}\right)\right]
 -\frac{c_{n1}}{k_{y}}\sqrt{2\varrho}e^{-i\pi/4}
 D_{\nu_n-1}\left(Q_{\gamma}\right).
\end{eqnarray}
The components of the spinor solution of the Dirac equation
\eqref{eqh1} in regions {\sf 2} and {\sf 4} can be obtained from
\eqref{hiii1} and \eqref{hiii2} with
$\varphi_{\gamma}^{+}(x)=\chi_{\gamma}^{+}+i\chi_{\gamma}^{-}$ and
$\varphi_{\gamma}^{-}(x)=\chi_{\gamma}^{+}-i\chi_{\gamma}^{-}$. Then, in
regions {\sf 2} and {\sf 4}
we have the eigenspinors
\begin{eqnarray}
 \Phi_{\sf j } &=& a_{\sf{j}-1}\left(%
\begin{array}{c}
 \eta^{+}_{\gamma}(x) \\
  \eta^{-}_{\gamma}(x) \\
\end{array}%
\right)e^{ik_{y}y}+a_{\sf j}\left(%
\begin{array}{c}
 \xi^{+}_{\gamma}(x) \\
 \xi^{-}_{\gamma}(x)\\
\end{array}%
\right)e^{ik_{y}y}
\end{eqnarray}
where
the functions $ \eta^{\pm}_{\gamma}(x)$ and $\xi^{\pm}_{\gamma}(x)$
are given by
\begin{eqnarray}
&& \eta^{\pm}_{\gamma}(x)=
 D_{\nu_{n}-1}\left(Q_{\gamma}\right)\mp
 \frac{1}{k_{y}}\sqrt{2\varrho}e^{i\pi/4}D_{\nu_{n}}\left(Q_{\gamma}\right)\\
&& \xi^{\pm}_{\gamma}(x)=
 \pm\frac{1}{k_{y}}\sqrt{2\varrho}e^{-i\pi/4}D_{-\nu_{n}+1}\left(-Q_{\gamma}^{*}\right)
  \pm
 \frac{1}{k_{y}}\left(-2i\epsilon_{0}\pm
 k_{y}-\gamma2i \varrho x\right)D_{-\nu_{n}}\left(-Q_{\gamma}^{*}\right).
\end{eqnarray}
More explicitly, it gives in region {\sf 2} 
\begin{eqnarray}
 \Phi_{\sf 2} &=& a_1\left(%
\begin{array}{c}
 \eta^{+}_{1}(x) \\
  \eta^{-}_{1}(x) \\
\end{array}%
\right)e^{ik_{y}y}+a_{2}\left(%
\begin{array}{c}
 \xi^{+}_{1}(x) \\
 \xi^{-}_{1}(x)\\
\end{array}%
\right)e^{ik_{y}y}
\end{eqnarray}
and region {\sf 4}
\begin{eqnarray}
 \Phi_{\sf 4} &=& a_3\left(%
\begin{array}{c}
 \eta^{+}_{-1}(x) \\
  \eta^{-}_{-1}(x) \\
\end{array}%
\right)e^{ik_{y}y}+a_{4}\left(%
\begin{array}{c}
 \xi^{+}_{-1}(x) \\
 \xi^{-}_{-1}(x)\\
\end{array}%
\right)e^{ik_{y}y}
\end{eqnarray}
where $\gamma=\pm 1$.

Solving the eigenvalue equation for
the Hamiltonian \eqref{eqh1} in region 3, we find the
following eigenspinor
\begin{eqnarray} \label{eq 7}
 && \Phi_{\sf 3}= b_1 \left(
            \begin{array}{c}
              {\alpha} \\
              {\beta z_{3}} \\
            \end{array}
          \right) e^{i(k_{3}x+k_{y}y)} +b_2 \left(
            \begin{array}{c}
              {\alpha} \\
              {-\beta z_{3}^{-1}} \\
            \end{array}
          \right) e^{i(-k_{3}x+k_{y}y)}\\
&& z_{3}
=s_{3}\frac{k_{3}+ik_{y}}{\sqrt{k_{3}^{2}+k_{y}^{2}}}
          \end{eqnarray}
where the parameters $\alpha$ and $\beta$ are defined by
\begin{equation} \label{eq 18i}
       {\alpha=\lp{1+\frac{\mu}{ \epsilon-v_{2}}}\gp}^{1/2}, \qquad
       {\beta=\lp{1-\frac{\mu}{ \epsilon-v_{2}}}\gp}^{1/2}
 \end{equation}
with the sign function
 $s_{3}=\mbox{sign}(\epsilon-v_{2})$.
The wave vector being
\beq
k_{3}= \sqrt{(\epsilon-v_{2})^{2}-\mu^{2}-{k_{y}}^{2}}.
\eeq
Finally the eigenspinor in region {\sf 5} can be expressed as
\begin{equation}\label{eq6}
 \Phi_{\sf 5}= t_{s,n} \left(
            \begin{array}{c}
              {1} \\
              {z_{1}} \\
            \end{array}
          \right) e^{i(k_{1}x+k_{y}y)}.
\end{equation}


The transmission and reflection coefficients $(r_{s,n},t_{s,n})$
can be determined using the boundary conditions, that is, continuity of the
eigenspinors at each interface. Next we will use the above
solutions to explicitly determine the corresponding coefficient.
Now, requiring the continuity of the spinor wavefunctions at each
junction interface gives rise to the following set of equations
\bqr
\label{eq11} &&\Phi_{\sf 1}(-d_2)= \Phi_{\sf
2}(-d_2)\\
&&\Phi_{\sf 2}(-d_1)= \Phi_{\sf 3}(-d_1)\\
&&\Phi_{\sf 3}(d_1)= \Phi_{\sf 4}(d_1)\\
&&\Phi_{\sf 4}(d_2)= \Phi_{\sf 5}(d_2).
\eqr
We prefer to express these relationships in terms of $2\times 2$
transfer matrices between different regions. For this, we write
\beq
\left(%
\begin{array}{c}
  a_{\sf j} \\
  b_{\sf j} \\
\end{array}%
\right)=M_{{\sf j}, {\sf j}+1}\left(%
\begin{array}{c}
  a_{{\sf j}+1} \\
  b_{{\sf j}+1} \\
\end{array}%
\right)
\eeq
where 
$M_{{\sf j}, {\sf j}+1}$ being the transfer matrices that couple the
wavefunction in the $j$-th region to the wavefunction in the
${\sf j} + 1$-th region. Finally, we obtain the full transfer matrix over the
whole double barrier which can be written, in an obvious notation,
as follows
\begin{equation}\label{systm1}
\left(%
\begin{array}{c}
  1 \\
  r_{s,n} \\
\end{array}%
\right)=\prod_{{\sf j}=1}^{4}M_{{\sf j}, {\sf j}+1}\left(%
\begin{array}{c}
  t_{s,n} \\
  0 \\
\end{array}%
\right)=M\left(%
\begin{array}{c}
  t_{s,n} \\
  0 \\
\end{array}%
\right)
\end{equation}
where the total transfer matrix $M=M_{12}\cdot M_{2
3}\cdot M_{34}\cdot M_{45}$ is given by
\begin{eqnarray}
&& M=\left(%
\begin{array}{cc}
  m_{11} & m_{12} \\
  m_{21} & m_{22} \\
\end{array}%
\right) 
\\
&& M_{12}=\left(%
\begin{array}{cc}
   e^{-\textbf{\emph{i}}k_{1} d_{2}} &e^{\textbf{\emph{i}}k_{1} d_{2}} \\
  z_{1}e^{-\textbf{\emph{i}}k_{1} d_{2}} & -z^{\ast}_{1} e^{\textbf{\emph{i}}k_{1} d_{2}} \\
\end{array}%
\right)^{-1}\left(%
\begin{array}{cc}
\eta_{1}^{+}(-d_2) &  \xi_{1}^{+}(-d_2)\\
 \eta_{1}^{-}(-d_2) & \xi_{1}^{-} (-d_2)\\
\end{array}%
\right)
\\
&& M_{23}=\left(%
\begin{array}{cc}
 \eta_{1}^{+}(-d_1) &  \xi_{1}^{+}(-d_1)\\
 \eta_{1}^{-}(-d_1) & \xi_{1}^{-} (-d_1)\\
\end{array}%
\right)^{-1}\left(%
\begin{array}{cc}
\alpha e^{-\textbf{\emph{i}}k_{3} d_{1}} &\alpha e^{\textbf{\emph{i}}k_{3} d_{1}} \\
  \beta z_{3}e^{-\textbf{\emph{i}}k_{3} d_{1}} & -\beta z^{\ast}_{3} e^{\textbf{\emph{i}}k_{3} d_{1}} \\
\end{array}%
\right)
\\
&& M_{34}=\left(%
\begin{array}{cc}
 \alpha e^{\textbf{\emph{i}}k_{3} d_{1}} &\alpha e^{-\textbf{\emph{i}}k_{3} d_{1}} \\
  \beta z_{3}e^{\textbf{\emph{i}}k_{3} d_{1}} & -\beta z^{\ast}_{3} e^{-\textbf{\emph{i}}k_{3} d_{1}} \\
\end{array}%
\right)^{-1}\left(%
\begin{array}{cc}
 \eta_{-1}^{+}(d_1) &  \xi_{-1}^{+}(d_1)\\
 \eta_{-1}^{-}(d_1) & \xi_{-1}^{-} (d_1)\\
\end{array}%
\right)
\\
&& M_{45}=\left(%
\begin{array}{cc}
 \eta_{-1}^{+}(d_2) &  \xi_{-1}^{+}(d_2)\\
 \eta_{-1}^{-}(d_2) & \xi_{-1}^{-} (d_2)\\
\end{array}%
\right)^{-1}\left(%
\begin{array}{cc}
  e^{\textbf{\emph{i}}k_{1} d_{2}} & e^{-\textbf{\emph{i}}k_{1} d_{2}} \\
  z_{1} e^{\textbf{\emph{i}}k_{1} d_{2}}  & -z_{1}^{\ast} e^{-\textbf{\emph{i}}k_{1} d_{2}}  \\
\end{array}%
\right).
\end{eqnarray}
These can be used 
 to evaluate the reflection and transmission amplitudes
\begin{equation}\label{eq 63}
 t_{s,n}=\frac{1}{m_{11}}, \qquad  r_{s,n}=\frac{m_{21}}{m_{11}}.
\end{equation}
Some symmetry relationship between the parabolic cylindric functions are worth mentioning. These are given by
\begin{equation}
\eta_{-1}^{\pm}(d_1)=\eta_{1}^{\pm}(-d_1),\qquad
\eta_{-1}^{\pm}(d_2)=\eta_{1}^{\pm}(-d_2)
\end{equation}
\begin{equation}
 \xi_{-1}^{\pm}(d_1)=\xi_{1}^{\pm}(-d_1),\qquad
 \xi_{-1}^{\pm}(d_2)=\xi_{1}^{\pm}(-d_2).
\end{equation}

We should point out at this stage that we were unfortunately
forced to adopt a somehow cumbersome notation for our wavefunction
parameters in different potential regions due to the relatively
large number of necessary subscripts and superscripts. Before
matching the eigenspinors at the boundaries, let us define the
following shorthand notation
\begin{equation}
\eta_{1}^{\pm}(-d_1)=\eta_{11}^{\pm},\qquad
 \eta_{1}^{\pm}(-d_2)=\eta_{12}^{\pm}
\end{equation}
\begin{equation}
 \xi_{1}^{\pm}(-d_1)=\xi_{11}^{\pm},\qquad
 \xi_{1}^{\pm}(-d_2)=\xi_{12}^{\pm}.
\end{equation}
At this level, we should determine the transmission amplitude $t_{s,n}$. After
some lengthy algebra, one can solve the linear system given in
\eqref{systm1} to obtain the transmission and reflection
amplitudes in closed form. As far as the transmission is concerned, we find
\begin{equation}
t_{s,n}=\frac{\alpha\beta e^{2i(k_{1}d_{2}+k_{3}d_{1})}
\left(1+z_{1}^{2}\right)\left(1+z_{3}^{2}\right)}{z_{3}\left(e^{4ik_{3}d_{1}}-1\right)\left(
\alpha^{2}\mathcal{Y}_{2}+\beta^{2}\mathcal{Y}_{1}\right)+\alpha\beta
\mathcal{Y}_{3}}\left(\xi_{11}^{+}\eta_{11}^{-}-\xi_{11}^{-}\eta_{11}^{+}\right)
\left(\xi_{12}^{-}\eta_{12}^{+}-\xi_{12}^{+}\eta_{12}^{-}\right)
 \end{equation}
where we have defined the following quantities
\begin{eqnarray}
 &&\mathcal{Y}_{1}=\left(\xi_{12}^{-}\eta_{11}^{+}-\xi_{11}^{+}\eta_{12}^{-}-
 \xi_{12}^{+}\eta_{11}^{+}z_{1}+\xi_{11}^{+}\eta_{12}^{+}z_{1}\right)\left( \xi_{11}^{+}\eta_{12}^{+}+
 \xi_{11}^{+}\eta_{12}^{-}z_{1}-\eta_{11}^{+}(\xi_{12}^{+}+\xi_{12}^{-}z_{1}\right)\\
 &&
 \mathcal{Y}_{2}=\left(\xi_{11}^{-}\eta_{12}^{+}-\xi_{11}^{-}\eta_{12}^{-}z_{1}-\eta_{11}^{-}(
 \xi_{12}^{+}+\xi_{12}^{+}z_{1}\right)
 \left( -\xi_{12}^{-}\eta_{11}^{-}+
 \xi_{12}^{+}\eta_{11}^{-}z_{1}-\xi_{11}^{-}(\eta_{12}^{-}+\eta_{12}^{+}z_{1}\right)\\
 &&
  \mathcal{Y}_{3}=\Gamma_{0}\left(1+z_{1}^{2}z_{3}^{2}\right)+\Gamma_{1}z_{1}\left(1-z_{3}\right)+\Gamma_{2}\left(z_{1}^{2}+z_{3}^{2}\right)
  +e^{4id_{1}k_{3}}\left(\Gamma_{3}+\Gamma_{4}\right)
\end{eqnarray}
as well as
 \begin{eqnarray}
 \Gamma_{0}&=&-\xi_{12}^{+}\xi_{12}^{-}\eta_{11}^{+}\eta_{11}^{-}
 +\xi_{11}^{+}\xi_{12}^{-}\eta_{11}^{-}\eta_{12}^{+}+
 \xi_{11}^{-}\xi_{12}^{+}\eta_{11}^{+}\eta_{12}^{-}-
 \xi_{11}^{+}\xi_{11}^{-}\eta_{12}^{+}\eta_{12}^{-}\\
 \Gamma_{1}&=&\left(\xi_{12}^{+}\right)^{2}\eta_{11}^{+}\eta_{11}^{-}
 -\left(\xi_{12}^{-}\right)^{2}\eta_{11}^{+}\eta_{11}^{-}-
 \xi_{11}^{-}\xi_{12}^{+}\eta_{11}^{+}\eta_{12}^{+}-
 \xi_{11}^{+}\xi_{12}^{+}\eta_{11}^{-}\eta_{12}^{+}\\\nonumber
 &&
 +\xi_{11}^{+}\xi_{11}^{-}\left(\eta_{12}^{+}\right)^{2}
 -\xi_{11}^{+}\xi_{11}^{-}\left(\eta_{12}^{-}\right)^{2}+
 \xi_{11}^{-}\xi_{12}^{-}\eta_{11}^{+}\eta_{12}^{-}+
 \xi_{11}^{+}\xi_{12}^{-}\eta_{11}^{-}\eta_{12}^{-}\\
 \Gamma_{2}&=&\xi_{12}^{+}\xi_{12}^{-}\eta_{11}^{+}\eta_{11}^{-}
 -\xi_{11}^{-}\xi_{12}^{-}\eta_{11}^{+}\eta_{12}^{+}-
 \xi_{11}^{+}\xi_{12}^{+}\eta_{11}^{-}\eta_{12}^{-}+
 \xi_{11}^{+}\xi_{11}^{-}\eta_{12}^{+}\eta_{12}^{-}\\
 \Gamma_{3}&=&\left(\xi_{12}^{+}\right)^{2}\eta_{11}^{+}\eta_{11}^{-}\left(z_{3}^{2}-1\right)
 -\xi_{11}^{-}\xi_{12}^{-}\eta_{11}^{+}\left[\eta_{12}^{+}\left(1+z_{1}^{2}z_{3}^{2}\right)-\eta_{12}^{-}z_{1}\left(z_{3}^{2}-1\right)\right]\\\nonumber
 &&
 +\xi_{11}^{-}\xi_{11}^{+}\left[\left(\eta_{12}^{+}\right)^{2}z_{1}
 -\left(\eta_{12}^{-}\right)^{2}z_{1}+\eta_{12}^{+}\eta_{12}^{-}\left(z_{1}^{2}-1\right)\left(z_{3}^{2}-1\right)\right]
 \\
 \Gamma_{4}&=&\xi_{12}^{-}\eta_{11}^{-}\left[-\xi_{12}^{-}\eta_{11}^{+}z_{1}\left(z_{3}^{2}-1\right)+
 \xi_{11}^{+}\left(\eta_{12}^{-}z_{0}\left(z_{3}^{2}-1\right)+\eta_{12}^{+}\left(z_{1}^{2}+z_{3}^{2}\right)\right)\right]\\\nonumber
 &&\xi_{12}^{+}\xi_{12}^{-}\eta_{11}^{+}\eta_{11}^{-}\left(z_{1}^{2}+1\right)\left(z_{1}^{3}-1\right)-\xi_{12}^{+}
 \xi_{11}^{+}\eta_{11}^{-}\left(\eta_{12}^{-}\left(1+z_{1}^{2}z_{3}^{2}\right)+\eta_{12}^{+}z_{1}\left(z_{1}^{3}-1\right)\right)\\\nonumber
 &&
 +\xi_{12}^{+}\xi_{11}^{-}\eta_{11}^{+}\left[\eta_{12}^{-}\left(z_{1}^{2}+z_{3}^{2}\right)+\eta_{12}^{+}z_{1}\left(1-z_{3}^{2}\right)\right].
\end{eqnarray}

Now we are ready for the computation of the reflection $R_{s,n}$
and transmission $T_{s,n}$ coefficients. For this purpose, we
introduce the associated current density $J$, which defines
$R_{s,n}$ and $T_{s,n}$ as
\begin{equation}
  T_{s,n}=\frac{ J_{\sf {tra}}}{ J_{\sf {inc}}},\qquad R_{s,n}=\frac{J_{\sf {ref}}}{ J_{\sf {inc}}}
\end{equation}
where $J_{\sf {\sf {inc}}}$, $J_{\sf {ref}}$ and $J_{\sf {\sf {tra}}}$
stand for the incident, reflected and transmitted components of
the current density, respectively. It is easy to show that the
current density $J$ reads as
\begin{equation}
J= e\upsilon_{F}\Phi_{\sf }^{\dagger}\sigma _{x}\Phi_{\sf }
\end{equation}
which gives the following results for the incident, reflected and
transmitted components
\begin{eqnarray}
&& J_{\sf {inc}}=  e\upsilon_{F}(\Phi_{\sf 1}^{+})^{\dagger}\sigma
_{x}\Phi_{\sf 1}^{+}
 \\
 && J_{\sf {ref}}= e\upsilon_{F} (\Phi_{\sf 1}^{-})^{\dagger}\sigma _{x}\Phi_{\sf 1}^{-}
 \\
 && J_{\sf {tra}}= e\upsilon_{F}\Phi_{\sf 5}^{\dagger}\sigma _{x}\Phi_{\sf 5}.
\end{eqnarray}
They allow us to express the transmission and reflection
probabilities as
\begin{equation}
  T_{s,n}=|t_{s,n}|^{2},
\qquad
  R_{s,n}=|r_{s,n}|^{2}.
\end{equation}

The above results will be investigated numerically for different
potential configurations to enable us to study the most important
features of our system. Obviously, we can
check that the probability conservation condition $T_{s,n}+R_{s,n}=1$ is well
satisfied. Let us consider Figure
\ref{fig1ab}a) where we show the transmission and reflection
probabilities versus the energy $\epsilon$. In the first energy interval $\epsilon \leq k_{y}$ we have no
transmission because it is a forbidden zone.
\begin{figure}[h!]
\centering
\includegraphics[width=8cm, height=5cm]{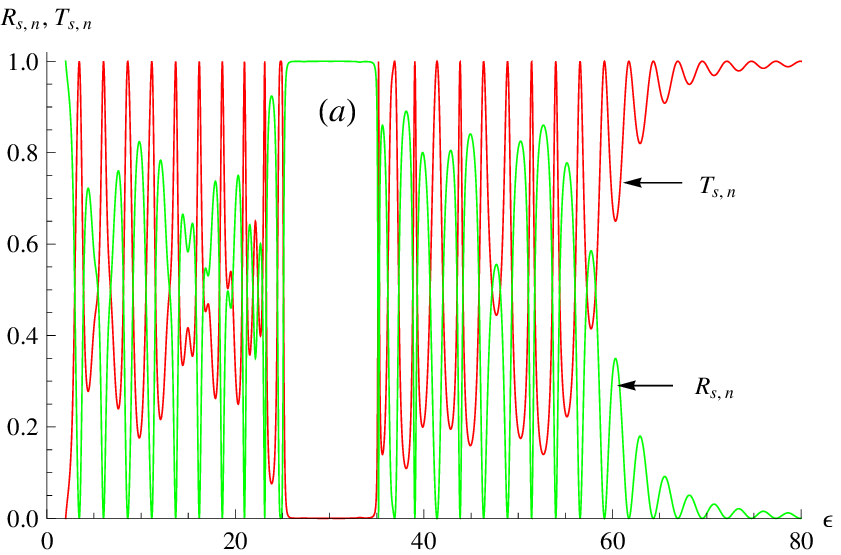}\ \ \ \
\includegraphics[width=8cm, height=5cm]{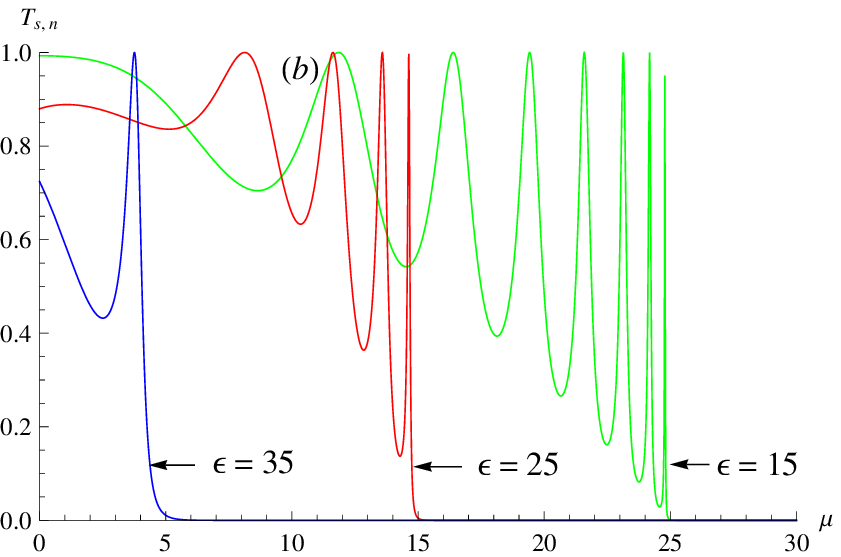}\\
 \caption{\sf{a) Transmission and reflection probabilities $(T_{s,n}, R_{s,n})$ as a function of energy
 $\epsilon$ with $d_{1}=0.6$, $d_{2}=2.5$, $\mu=4$,
 $k_{y}=2$,
 $v_{1}=60$ and $v_{2}=30$. b)Transmission  probability $T_{s,n}$ as a function of
 energy gap $\mu$ with $d_{1}=0.5$, $d_{2}=1.5$, $\epsilon=\{15, 25, 35\}$,
 $k_{y}=1$,
 $v_{1}=50$ and $v_{2}=40$
.}}\lb{fig1ab}
\end{figure}
However, for in second energy intervals $k_{y} \leq \epsilon \leq
v_{2}-k_{y}-\frac{\mu}{2}$ and $v_{2}+k_{y}+\frac{\mu}{2}\leq\epsilon\leq v_{1}$, we observe
resonance oscillations due to the Klein regime. We have no transmission
(like a windows) when $v_{2}-k_{y}-\frac{\mu}{2}\leq\epsilon\leq
v_{2}+k_{y}+\frac{\mu}{2}$. Finally in the interval where
$\epsilon > v_{1}$, there exist usual high energy
oscillations, which asymptotically saturates at high energy. Note
that  \eqref{eq 18i} implies that for certain energy gap $\mu$,
there is no transmission. In fact, under the condition
\begin{equation}
\mu>|v_{2}-\epsilon|
\end{equation}
every incoming wave is reflected. In Figure \ref{fig1ab}b)  we see that the transmission vanishes
for values of $\epsilon$ below the critical value
$\mu=|v_{2}-\epsilon|$.\\

\begin{figure}[h!]
\centering
\includegraphics[width=8cm, height=5cm]{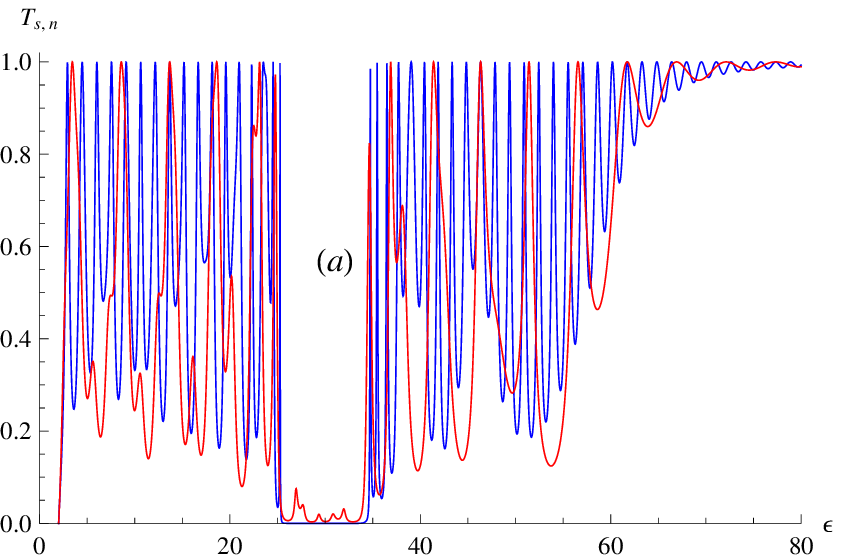}\ \ \ \
\includegraphics[width=8cm, height=5cm]{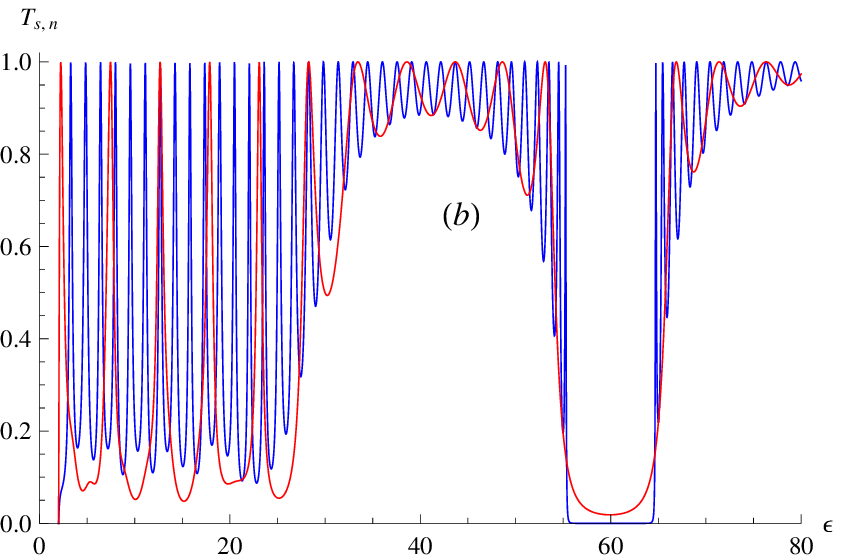}\\
 \caption{\sf{(Color online) Transmission probability for the static barrier $T_{s,n}$ as a function of energy
 $\epsilon$ with  $d_{1}=0.3$ color red, $d_{1}=1$, $d_{2}=2.5$, $\mu=4$ and $k_{y}=2$. a) the parameters:
 $v_{1}=60$ , $v_{2}=30$. b) $v_{1}=30$, $v_{2}=60$.}}\lb{figgi3}
\end{figure}
Figure \ref{figgi3} presents the transmission $T_{s,n}$ as a
function of incident electron energy $\epsilon$ for the Dirac
fermion scattered by a double triangular barriers with
$d_{2}=2.5$, $\mu=4$, $k_{y}=2$ and two values of
barrier height $d_{1}=\{0.3, 1\}$. We consider in Figure
\ref{figgi3}a) the parameters:
 $v_{1}=2 v_{2}=60$, the results show that as long as the well width
$d_{1}$ increases the transmission resonance shifts and the width
of the resonances increases between $k_{y} \leq \epsilon \leq
v_{2}-k_{y}-\frac{\mu}{2}$ and
$v_{2} + k_{y} + \frac{\mu}{2}\leq\epsilon\leq v_{1}$. In Figure
\ref{figgi3}b) we consider the parameters
$v_{1}=\frac{v_{2}}{2}=30$ for the Dirac fermion scattered by a
double barrier triangular potential where we distinguish five  different zones.
\begin{itemize}
 \item 
The first is a forbidden zone where 0$ \leq \epsilon \leq k_{y}$. 
\item The
second zone $k_{y} \leq \epsilon \leq v_{1}$ is the upper Klein
energy zone with transmission resonances.
\item The third zone contains
oscillations.
\item The fourth one
$v_{2}-k_{y}-\frac{\mu}{2}\leq\epsilon\leq
v_{2}+k_{y}+\frac{\mu}{2}$ is a window where the transmission is
zero, the wavefunction is damped and transmission decays
exponentially. 
\item The fifth zone  $\epsilon\geq
v_{2}+k_{y}+\frac{\mu}{2}$ contains oscillations, the transmission
converges to unity at high energies similarly to the
non-relativistic result.
\end{itemize}

\begin{figure}[h!]
\centering
\includegraphics[width=8cm, height=5cm]{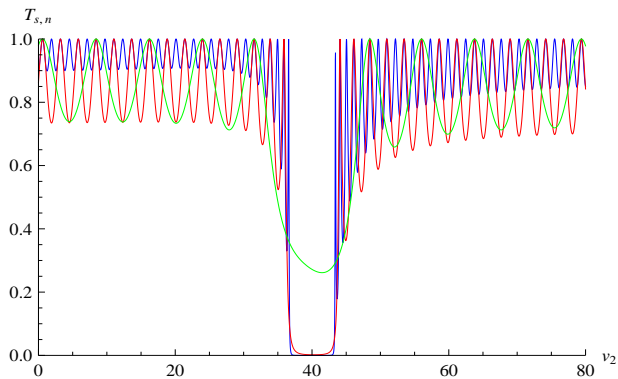}\\
 \caption{\sf{(Color online) Transmission probability for the static barrier $T_{s,n}$ as a function of
 energy potential
 $v_{2}$ with $d_{1}=0.2$ color red, $d_{1}=0.6$ color green, $d_{1}=1.2$ color blue, $d_{2}=2$, $\mu=3$,
 $k_{y}=1$,
 $\epsilon=40$ and $v_{1}=60$.}}\lb{figg3}
\end{figure}

\begin{figure}[h!]
\centering
\includegraphics[width=8cm, height=5cm]{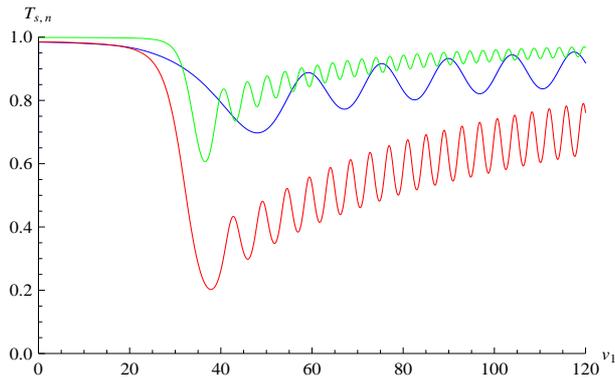}\\
 \caption{\sf{(Color online) Transmission probability  for the static barrier $T_{s,n}$ as a function of
 energy potential
 $v_{1}$ with $d_{1}=0.7$ color red, $d_{1}=2$ color blue, $d_{1}=0.05$ color green, $d_{2}=2.5$, $\mu=4$, $k_{y}=2$,
 $\epsilon=30$ and $v_{2}=60$.
.}}\lb{figg4}
\end{figure}
We represent in Figure \ref{figg3} the transmission versus potential energy
 $v_{2}$. It is clear that the two transmission curves are
symmetric with respect to the point $v_{2} = \epsilon$.
While an increase in the value $d_{1}$ widens the bowl width.
Figure \ref{figg4} presents the transmission probability  for a
static barrier $T_{s,n}$  as function of the
strength of the applied voltage $v_{1}$. The transmission is
observed for small values of $v_{1}$ less than the energy of the
incident fermion. It then decreases sharply for $v_{1} > \epsilon
-(2k_{y}+\mu)$ until it reaches a relative minimum and then begins
to increase in an oscillatory manner.

\section{Magnetic double barrier}
\begin{figure}[h]
  \centering
  \includegraphics[width=14cm, height=5cm ]{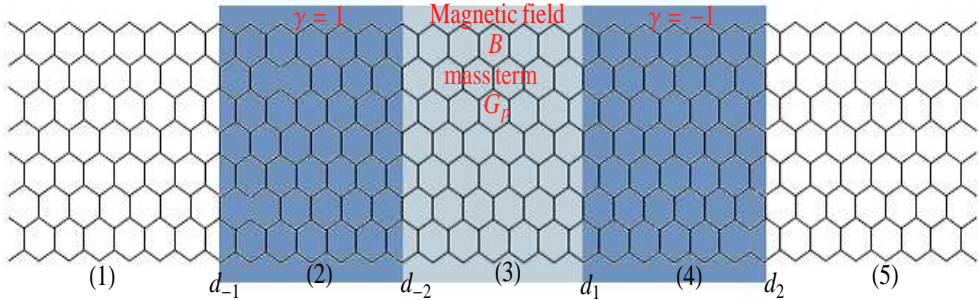}\\
  \caption{\sf Schematic diagram for the monolayer graphene double barrier.}\label{fig.1}
\end{figure}

Consider a two-dimensional system of Dirac fermions forming a graphene
sheet. This sheet is subject to a double
barrier potential in addition to a mass term and an externally
applied magnetic field as shown in Figure \ref{fig.1}. Particles
and antiparticles moving respectively in the positive and negative
energy regions with the tangential component of the wave vector
along the $x$-direction have translation invariance in the
$y$-direction. 
A uniform perpendicular
magnetic field is applied, along the $z$-direction and confined to the well
region between the two barriers. It is defined by
\begin{equation}\label{eq04}
B(x,y)=B\Theta(d_{1}^{2}-x^{2})
\end{equation}
where $B$ is the strength of the magnetic field within the strip
located in the region $|x|< d_{1}$ and $B=0$ otherwise, $\Theta$ is
the Heaviside step function. Choosing the Landau gauge and
imposing continuity of the vector potential at the boundary to
avoid unphysical effects, we end up with the following vector
potential
\begin{equation}
\qquad A_{y}(x)=A_{j}=\frac{c}{e}\times\left\{%
\begin{array}{ll}
  -\frac{1}{l_{B}^{2}}d_{1}, & \hbox{$x<-d_{2}$} \\
    \frac{1}{l_{B}^{2}}x, & \hbox{$\mid x\mid<d_{1}$} \\
     \frac{1}{l_{B}^{2}}d_{1}, & \hbox{$x\geq d_{2}$} \\
\end{array}%
\right.
\end{equation}
with the magnetic length is $l_{B}=\sqrt{1/B}$ in the unit
system ($\hbar=c=e=1$). 
The system contains five regions denoted ${\sf j=1,2,3,4,5}$. The
left region (${\sf j=1}$) describes the incident electron beam
with the energy $E=v_{F}\epsilon$ at an incident angle
$\phi_{1}$  where $v_{F}$ is the Fermi velocity. The extreme right region
(${\sf j=5}$) describes the transmitted electron beam at an
angle $\phi_{5}$. The Hamiltonian for one-pseudospin component
describing our system reads as
\begin{equation} \lb{equm1}
H_{m}=v_{F}
{\boldsymbol{\sigma}}\cdot\left(\textbf{p}+\frac{e}{c}\textbf{A}\right)+
V(x){\mathbb I}_{2}+G_p\Theta\left(d_{1}^{2}-x^{2}\right)\sigma_{z}
\end{equation}
{To proceed further,} we need to find the solutions of
the corresponding Dirac equation and their associated energy
spectrum.

\subsection{Energy spectrum solutions}
We are set to determine the eigenvalues and eigenspinors of the Hamiltonian
 $H_{m}$. Indeed, the Dirac Hamiltonian describing regions 1 and 5, is
obtained from \eqref{equm1} as
\begin{equation}
H_{m}=\left(%
\begin{array}{cc}
  0 & \upsilon_{F}\left(p_{x{\sf j}}-i\left(p_{y}+\frac{e}{c}A_{\sf j}\right)\right) \\
  \upsilon_{F}\left(p_{x{\sf j}}+i\left(p_{y}+\frac{e}{c}A_{\sf j}\right)\right) & 0 \\
\end{array}%
\right).
\end{equation}
The corresponding time independent Dirac equation for the spinor 
$\psi_{\sf j}(x,y)= (\varphi_{\sf j}^{+}, \varphi_{\sf j}^{-})^{T}$ at energy
$E=\upsilon_{F}\epsilon$ is given by
\begin{equation}
H_{m}\left(%
\begin{array}{c}
  \varphi_{\sf j}^{+} \\
  \varphi_{\sf j}^{-} \\
\end{array}%
\right)=\epsilon\left(%
\begin{array}{c}
  \varphi_{\sf j}^{+} \\
  \varphi_{\sf j}^{-} \\
\end{array}%
\right).
\end{equation}
This eigenproblem can be written as two linear differential
equations of the from
\begin{eqnarray}
     &&p_{x{\sf j}}-i\left(p_{y}+\frac{e}{c}A_{\sf j}\right)\varphi_{\sf j}^{-}=\epsilon\varphi_{\sf j}^{+}  \\
     && p_{x{\sf j}}+i\left(p_{y}+\frac{e}{c}A_{\sf j}\right)\varphi_{\sf j}^{+}=\epsilon\varphi_{\sf j}^{-}  
\end{eqnarray}
which gives the energy eigenvalue 
\begin{equation}
\epsilon=s_{\sf j} \sqrt{p_{x{\sf j}}^{2}+\left(p_{y}+\frac{e}{c}A_{\sf j}\right)}
\end{equation}
where $s_{\sf j}=\mbox{sign}(\epsilon)$. This  implies
\begin{equation}
p_{x{\sf j}}=\sqrt{\epsilon ^{2}-\left(p_{y}+\frac{e}{c}A_{\sf j}\right)^{2}}
\end{equation}
with incoming momentum ${\boldsymbol{p_{\sf j}}}=(p_{x{\sf j}}, p_{y})$ and
${\boldsymbol{r}}=(x, y)$. The incoming wave function is
\begin{eqnarray}
&& \psi_{in}=\frac{1}{\sqrt{2}}\left(
\begin{array}{c}
1 \\
 z_{p_{x{\sf j}}}\end{array}\right)e^{\textbf{\emph{i}}{\boldsymbol{p_{\sf j}}}{\boldsymbol{r}}}\\
&& z_{p_{x{\sf j}}}=z_{\sf j}=s_{\sf j}\frac{p_{x{\sf j}}
+i(p_{y}+\frac{e}{c}A_{\sf j})}{\sqrt{(p_{x{\sf j}})^{2}
+(p_{y}+\frac{e}{c}A_{\sf j})^{2}}}=s_{\sf j}
e^{\textbf{\emph{i}}\phi_{\sf j}}
\end{eqnarray}
where  $s_{0}=\mbox{sgn}(\epsilon)$ and  $\phi_{\sf j}=\arctan\left(\frac{p_{y}-\frac{e}{c}A_{\sf j}}{p_{x{\sf j}}}\right)$ is the angle that
the incident electrons make with the {$x$-direction}, $p_{x1}$ and
$p_{y}$ are the $x$ and $y$-components of the electron wave
vector, respectively. The eigenspinors are given by
\begin{eqnarray}
&& \psi_{\sf j}^{+}=\frac{1}{\sqrt{2}}\left(
\begin{array}{c}
1 \\
 z_{\sf j}\end{array}\right)e^{\textbf{\emph{i}}(p_{x{\sf j}} x +p_{y} y)}\\
&&
\psi_{\sf j}^{-}=\frac{1}{\sqrt{2}}\left(
\begin{array}{c}
1 \\
 -z^{*}_{\sf j}\end{array}\right)e^{\textbf{\emph{i}}(-p_{x{\sf j}} x +p_{y} y)}.
\end{eqnarray}
It is straightforward to solve the tunneling problem for Dirac
fermions. We assume that the incident  wave propagates at the angle
$\phi_{1}$ with respect to the {$x$-direction} and write  the
components, of the Dirac spinor $\varphi_{\sf j}^{+}$ and
$\varphi_{\sf j}^{-}$, for each
region, in the following form

$\star$ For $x<-d_{2}$ (region 1):
\begin{eqnarray}
&& \epsilon=
\left[p_{x1}^{2}+\left(p_{y}-\frac{1}{l_{B}^{2}}d_{1}\right)^{2}\right]^{\frac{1}{2}}\\
&& \psi_{1}=\frac{1}{\sqrt{2}}\left(
\begin{array}{c}
1 \\
 z_{1}\end{array}\right)e^{\textbf{\emph{i}}(p_{x1} x +p_{y} y)}+r_{m}\frac{1}{\sqrt{2}}\left(
\begin{array}{c}
1 \\
 -z^{*}_{1}\end{array}\right)e^{\textbf{\emph{i}}(-p_{1x} x +p_{y}
 y)}\\
&& z_{1}=s_{1}\frac{p_{x1}
+i\left[p_{y}-\frac{1}{l_{B}^{2}}d_{1}\right]}{\sqrt{p_{x1}^{2}
+\left[p_{y}-\frac{1}{l_{B}^{2}}d_{1}\right]^{2}}}.
\end{eqnarray}

$\star$ In the barrier $x
> d_{2}$ (region 5)
\begin{eqnarray}
&& \epsilon=\left[p_{x5}^{2}+\left(p_{y}+\frac{1}{l_{B}^{2}}d_{1}\right)^{2}\right]^{\frac{1}{2}}\\
&& \psi_{5}=\frac{1}{\sqrt{2}}t_{m}\left(
\begin{array}{c}
1 \\
 z_{5}\end{array}\right)e^{\textbf{\emph{i}}(p_{x5} x +p_{y} y)}\\
&& z_{5}=s_{5}\frac{p_{x5}
+i\left[p_{y}+\frac{1}{l_{B}^{2}}d_{1}\right]}{\sqrt{p_{x1}^{2}
+\left[p_{y}+\frac{1}{l_{B}^{2}}d_{1}\right]^{2}}}.
\end{eqnarray}

$\star$ In region {\sf 2} and {\sf 4} ($d_{1}<|x|<d_{2}$):
The general solution can be expressed in terms of the parabolic
cylinder function \cite{Abramowitz, Gonzalez, HBahlouli} as
\begin{equation}\lb{hii1}
 \chi_{\gamma}^{+}=c_{1}
 D_{\nu_\gamma-1}\left(Q_{\gamma}\right)+c_{2}
 D_{-\nu_\gamma}\left(-Q^{*}_{\gamma}\right)
\end{equation}
where
$\nu_{\gamma}=\frac{i}{2\varrho}\left(k_{y}-\gamma\frac{d_{1}}{l_{B}^{2}}\right)^{2}$,
$\epsilon_{0}=\epsilon-v_{1}$ and $
Q_{\gamma}(x)=\sqrt{\frac{2}{\varrho}}e^{i\pi/4}\left(\gamma
\varrho x+\epsilon_{0}\right) $, $c_{1}$ and $c_{2}$ are
constants and gives the other component
\begin{eqnarray}\lb{hii2}
\chi_{\gamma}^{-}&=&-c_{2}\frac{1}{k_{y}-\gamma\frac{d_{1}}{l_{B}^{2}}}\left[
2(\epsilon_{0}+\gamma \varrho x)
 D_{-\nu_\gamma}\left(-Q^{*}_{\gamma}\right)
+
 \sqrt{2\varrho}e^{i\pi/4}D_{-\nu_\gamma+1}\left(-Q^{*}_{\gamma}\right)\right]\nonumber\\
 &&
 -\frac{c_{1}}{k_{y}-\gamma\frac{d_{1}}{l_{B}^{2}}}\sqrt{2\varrho}e^{-i\pi/4}
 D_{\nu_\gamma-1}\left(Q_{\gamma}\right)
\end{eqnarray}
The components of the spinor solution of the Dirac equation
\eqref{eqh1} in region {\sf 2} and {\sf 4} can be obtained from
\eqref{hii1} and \eqref{hii2} with
$\varphi_{\gamma}^{+}(x)=\chi_{\gamma}^{+}+i\chi_{\gamma}^{-}$ and
$\varphi_{\gamma}^{-}(x)=\chi_{\gamma}^{+}-i\chi_{\gamma}^{-}$. We
have the eigenspinor
\begin{eqnarray}
 \psi_{\sf j} &=& a_{{\sf j}-1}\left(%
\begin{array}{c}
 u^{+}_{\gamma}(x) \\
  u^{-}_{\gamma}(x) \\
\end{array}%
\right)e^{ik_{y}y}+a_{\sf j}\left(%
\begin{array}{c}
 v^{+}_{\gamma}(x) \\
 v^{-}_{\gamma}(x)\\
\end{array}%
\right)e^{ik_{y}y}
\end{eqnarray}
where ${\sf j=2, 4}$ and $\gamma=\pm 1$, the function
$u^{\pm}_{\gamma}(x)$ and $v^{\pm}_{\gamma}(x)$ are given by
\begin{eqnarray}
u^{\pm}_{\gamma}(x)&=&
 D_{\nu_{\gamma}-1}\left(Q_{\gamma}\right)\mp
 \frac{1}{k_{y}-\gamma\frac{d_{1}}{l_{B}^{2}}}\sqrt{2\varrho}e^{i\pi/4}D_{\nu_{\gamma}}\left(Q_{\gamma}\right)
\\
v^{\pm}_{\gamma}(x)&=&
 \pm\frac{1}{k_{y}-\gamma\frac{d_{1}}{l_{B}^{2}}}\sqrt{2\varrho}e^{-i\pi/4}D_{-\nu_{\gamma}+1}\left(-Q_{\gamma}^{*}\right)\nonumber\\
 &&
  \pm
 \frac{1}{k_{y}-\gamma\frac{d_{1}}{l_{B}^{2}}}\left(-2i\epsilon_{0}\pm
 \left(k_{y}-\gamma\frac{d_{1}}{l_{B}^{2}}\right)-\gamma2i \varrho x\right)D_{-\nu_{\gamma}}\left(-Q_{\gamma}^{*}\right).
\end{eqnarray}
In region {\sf 2}: 
\begin{eqnarray}
 \psi_{\sf 2} &=& a_1\left(%
\begin{array}{c}
 u^{+}_{1}(x) \\
  u^{-}_{1}(x) \\
\end{array}%
\right)e^{ik_{y}y}+a_{2}\left(%
\begin{array}{c}
 v^{+}_{1}(x) \\
 v^{-}_{1}(x)\\
\end{array}%
\right)e^{ik_{y}y}
\end{eqnarray}
In region {\sf 4}:
\begin{eqnarray}
 \psi_{\sf 4} &=& a_3\left(%
\begin{array}{c}
 u^{+}_{-1}(x) \\
  u^{-}_{-1}(x) \\
\end{array}%
\right)e^{ik_{y}y}+a_{4}\left(%
\begin{array}{c}
 v^{+}_{-1}(x) \\
 v^{-}_{-1}(x)\\
\end{array}%
\right)e^{ik_{y}y}
\end{eqnarray}

$\star$ In the region $|x|\leq d_{1}$:
From the nature of the system under consideration, we write the
Hamiltonian corresponding to region ${\sf 3}$ in matrix form as
\begin{equation}\label{eq 20}
H_m=v_{F}\left(%
\begin{array}{cc}
  \frac{V_{2}}{v_{F}}+\frac{G_{p}}{v_{F}} & -i\frac{\sqrt{2}}
  {l_{B}}\left(\frac{l_{B}}{\sqrt{2}}\left(\partial_{x}-i\partial_{y}+\frac{e}{c}A_{3}\right)\right)\\
 i\frac{\sqrt{2}}{l_{B}}\left(\frac{l_{B}}{\sqrt{2}}\left(-\partial_{x}-i\partial_{y}
 +\frac{e}{c}A_{3}\right)\right)  &  \frac{V_{2}}{v_{F}}-\frac{G_{p}}{v_{F}}\\
\end{array}%
\right)
\end{equation}
Note that, the energy gap $G_{p}$ behaves like a mass term in Dirac equation.
Certainly this will affect the above results and lead to
interesting consequences on the transport properties of our
system. We determine the eigenvalues and eigenspinors of the
Hamiltonian $H_m$ by considering the time independent equation for the spinor
$\psi_{3}(x, y)=(\psi_{3}^{+}, \psi_{3}^{-})^{T}$ using the fact that the
transverse momentum $p_{y}$ is conserved, we can write the wave
function
$\psi_{3}(x, y)=e^{ip_{y}y} \varphi_{3}(x)$
with $\varphi_{3}(x)= (\varphi_{3}^+, \varphi_{3}^-)^{T}$, the energy being defined by
$E=\upsilon_{F}\epsilon$ leads to
\begin{equation}\label{eq 23}
H_{m}\left(%
\begin{array}{c}
  \varphi_{3}^+ \\
  \varphi_{3}^-\\
\end{array}%
\right)=\epsilon\left(%
\begin{array}{c}
  \varphi_{3}^+\\
  \varphi_{3}^-\\
\end{array}%
\right)
\end{equation}
At this stage, it is convenient to introduce the annihilation and
creation operators. They can be defined as
\begin{eqnarray}
a=\frac{l_{B}}{\sqrt{2}}\left(\partial_{x}+k_{y}+\frac{e}{c}A_{3}\right),
\qquad
a^{\dagger}=\frac{l_{B}}{\sqrt{2}}\left(-\partial_{x}+k_{y}+\frac{e}{c}A_{3}\right)
\end{eqnarray}
which obey the canonical commutation relations $[a,
a^{\dagger}]={\mathbb I}$. Rescaling our energies $G_{p}=\upsilon_{F}\mu$
and $V_{2}=\upsilon_{F}v_{2}$, \eqref{eq 23} can be
written in terms of $a$ and $a^{\dagger}$ as
\begin{equation}
 \left(%
\begin{array}{cc}
  v_{2}+\mu & -i\frac{\sqrt{2}}{l_{B}}a \\
  +i\frac{\sqrt{2}}{l_{B}}a^{\dagger}  &  v_{2}-\mu \\
\end{array}%
\right)\left(%
\begin{array}{c}
  \varphi_{3}^+ \\
  \varphi_{3}^- \\
\end{array}%
\right)=\epsilon\left(%
\begin{array}{c}
  \varphi_{3}^+ \\
  \varphi_{3}^- \\
\end{array}%
\right)
\end{equation}
which gives
\begin{eqnarray}\label{eq 25}
  &&(v_{2}+\mu)\varphi_{3}^{+}-i\frac{\sqrt{2}}{l_{B}}a\varphi_{3}^-=\epsilon\varphi_{3}^+\\
&&\label{eq 26}
  i\frac{\sqrt{2}}{l_{B}}a^{\dagger}\varphi_{3}^+ +
  (v_{2}-\mu)\varphi_{3}^{-}=\epsilon\varphi_{3}^-.
\end{eqnarray}
Injecting \eqref{eq 26} in \eqref{eq 25}, we obtain a
differential equation of second order for
 $\varphi_{3}^{+}$
\begin{equation}
\left[(\epsilon-v_{2})^{2}-\mu^{2}\right]\varphi_{3}^{+}=\frac{2}{l_{B}^{2}}a
a^{\dagger}\varphi_{3}^{+}.
\end{equation}
It is clear that $\varphi_{3}^{+}$ is an eigenstate of the number
operator $\widehat{N}=a^{\dagger}a$ and therefore we identify
$\varphi_{3}^{+}$ with the eigenstates of the harmonic oscillator
$|n-1\rangle$, namely
\begin{equation}
 \varphi_{3}^{+} \sim \mid n-1\rangle
\end{equation}
which is equivalent to
 \begin{equation} 
 \left[(\epsilon-v_{2})^{2}-\mu^{2}\right] 
 \mid n-1\rangle=\frac{2}{l_{B}^{2}}n\mid n-1\rangle
\end{equation}
and the associated energy spectrum is
\begin{equation}
\epsilon-v_{2}=s_{3} \epsilon_{n}=s_{3}\frac{1}{l_{B}}\sqrt{(\mu
l_{B})^{2}+2n}
\end{equation}
where we have set $\epsilon_{n}=s_{3}(\epsilon-v_{2})$ and
$s_{3}=\mbox{sign}(\epsilon_{n}-v_{2})$ correspond to positive and
negative energy solutions. For this reason we write the eigenvalues
as
\begin{equation}
\epsilon=v_{2}+s_{3}\frac{1}{l_{B}}\sqrt{(\mu l_{B})^{2}+2n}
\end{equation}
The second eigenspinor component then can be obtained from
\begin{equation}
\varphi_{3}^{-}=\frac{i\sqrt{2}a^{\dagger}}{(\epsilon-v_{2})l_{B}+\mu
l_{B}}\mid n-1\rangle=
       \frac{i\sqrt{2n}}{(\epsilon-v_{2})l_{B}+\mu l_{B}} \mid n\rangle
\end{equation}
where $ \sqrt{2n}=\sqrt{(\epsilon_n l_{B})^{2}-(\mu
l_{B})^{2}}$. We find
\begin{equation}
\varphi_{3}^{-}=s_{3}i\sqrt{\frac{\epsilon_n l_{B}-s_{3} \mu
l_{B}}{\epsilon_n l_{B}+s_{3} \mu l_{B}}} \mid n\rangle
\end{equation}
After normalization we arrive at the following expression for the positive and negative energy
eigenstates
\begin{equation}
\varphi_{3}=\frac{1}{\sqrt{2}}\left(%
\begin{array}{c}
  \sqrt{\frac{\epsilon_n l_{B}+s_{3} \mu l_{B}}{\epsilon_n l_{B}}} \mid n-1\rangle \\
  s_{3} i\sqrt{\frac{\epsilon_n l_{B}-s_{3} \mu l_{B}}{\epsilon_n l_{B}}} \mid n\rangle \\
\end{array}%
\right)
\end{equation}
Introducing the parabolic cylinder functions
$D_{n}(x)=2^{-\frac{n}{2}}e^{-\frac{x^{2}}{4}}H_{n}\left(\frac{x}{\sqrt{2}}\right)$ to express
the solution in region {\sf 3} as
\begin{equation} \psi_{\sf
3}(x,y)=b_{1}\psi_{3}^{+}+b_{2}\psi_{3}^{-}
\end{equation} 
with the two components
\begin{equation}
\psi_{3}^{\pm}(x, y)=\frac{1}{\sqrt{2}}\left(%
\begin{array}{c}
 \sqrt{\frac{\epsilon_n l_{B}+s_{3} \mu l_{B}}{\epsilon_n l_{B}}}
 D_{\left(\left(\epsilon_n l_{B}\right)^{2}-(\mu l_{B})^{2} \right)/2-1}
 \left(\pm \sqrt{2}\left(\frac{x}{l_{B}}+k_{y}l_{B}\right)\right) \\
  \pm i\frac{s_{3}\sqrt{2}}{\sqrt{\epsilon_n l_{B}\left(\epsilon_n l_{B}+s_{3} \mu l_{B}\right)}}
  D_{\left(\left(\epsilon_n l_{B}\right)^{2}-\left(\mu l_{B}\right)^{2}\right)/2}
 \left(\pm \sqrt{2}\left(\frac{x}{l_{B}}+k_{y}l_{B}\right)\right) \\
\end{array}%
\right)e^{ik_{y}y}
\end{equation}

As usual the coefficients $(a_1,a_2,a_3,a_4,b_1,b_2,r,t)$ can be
determined using the boundary conditions, continuity of the
eigenspinors at each interface.

\subsection{ Transmission and reflection amplitudes}
 We will now study some
interesting features of our system in terms of the
corresponding transmission probability. Before doing so, let us
simplify our writing using the following shorthand notation
\begin{eqnarray}
&&\vartheta_{\tau1}^{\pm}=D_{\left[(\epsilon_{n}l_{B})^{2}-(\mu
l_{B})^{2}\right]/2-1}
 \left[\pm \sqrt{2}\left(\frac{\tau d_{1}}{l_{B}}+k_{y}l_{B}\right)\right]\\
&& \zeta_{\tau1}^{\pm}= D_{\left[(\epsilon_{n}l_{B})^{2}-(\mu
l_{B})^{2}\right]/2}
  \left[\pm \sqrt{2}\left(\frac{\tau d_{1}}{l_{B}}+k_{y}l_{B}\right)\right]\\
  && f_{1}^{\pm}=\sqrt{\frac{\epsilon_{n}\pm
\mu}{\epsilon_{n}}}, \qquad
f_{2}^{\pm}=\frac{\sqrt{2/l_{B}^{2}}}{\sqrt{\epsilon_{n}(\epsilon_{n}\pm
\mu)}}\\
&& u^{\pm}_{\gamma}(\tau d_{1})=u^{\pm}_{\gamma, \tau1},\qquad
u^{\pm}_{\gamma}(\tau d_{2})=u^{\pm}_{\gamma, \tau 2}\\
&& v^{\pm}_{\gamma}(\tau d_{1})=v^{\pm}_{\gamma, \tau 1},\qquad
v^{\pm}_{\gamma}(\tau d_{2})=v^{\pm}_{\gamma, \tau 2}
\end{eqnarray}
where $\tau=\pm$. Dirac equation requires the following set
of continuity equations
\bqr \label{eq11} &&\psi_{\sf 1}(-d_2)= \psi_{\sf
2}(-d_2)\\
&&\psi_{\sf 2}(-d_1)= \psi_{\sf 3}(-d_1)\\
&&\psi_{\sf 3}(d_1)= \psi_{\sf 4}(d_1)\\
&&\psi_{\sf 4}(d_2)= \psi_{\sf 5}(d_2) \eqr
That is requiring the continuity of the spinor wave functions at each
junction interface give rise to the above set of equations. We prefer to
express these relationships in terms of $2\times 2$ transfer
matrices between {\sf j}-th and {\sf j}+1-th regions, $\mathcal{M}_{{\sf j},{\sf j}+1}$, we
obtain the full transfer matrix over the whole double barrier
which can be written, in an obvious notation, as follows
\begin{equation}\label{syst1}
\left(%
\begin{array}{c}
  1 \\
  r_{m} \\
\end{array}%
\right)=\prod_{{\sf j}=1}^{4}\mathcal{M}_{{\sf j},{\sf j}+1}\left(%
\begin{array}{c}
  t_{m} \\
  0 \\
\end{array}%
\right)=\mathcal{M}\left(%
\begin{array}{c}
  t_{m} \\
  0 \\
\end{array}%
\right)
\end{equation}
where the total transfer matrix $\mathcal{M}=\mathcal{M}_{12}\cdot \mathcal{M}_{2
3}\cdot \mathcal{M}_{34}\cdot\mathcal{ M}_{45}$ are transfer
matrices that couple the wave function in the ${\sf j}$-th region to the
wave function in the ${\sf j} + 1$-th region. These are given explicitly by
\begin{eqnarray}
&& \mathcal{M}=\left(%
\begin{array}{cc}
  \tilde{m}_{11} & \tilde{m}_{12} \\
  \tilde{m}_{21} & \tilde{m}_{22} \\
\end{array}%
\right)\\
&& \mathcal{M}_{12}=\left(%
\begin{array}{cc}
   e^{-\textbf{\emph{i}}p_{x1} d_{2}} &e^{\textbf{\emph{i}}p_{x1} d_{2}} \\
  z_{1}e^{-\textbf{\emph{i}}p_{x1} d_{2}} & -z^{\ast}_{1} e^{\textbf{\emph{i}}p_{x1} d_{2}} \\
\end{array}%
\right)^{-1}\left(%
\begin{array}{cc}
u_{1,-2}^{+} &  v_{1,-2}^{+}\\
 u_{1,-2}^{-} & v_{1,-2}^{-}\\
\end{array}%
\right)\\
&& \mathcal{M}_{23}=\left(%
\begin{array}{cc}
 u_{1,-1}^{+} &  v_{1,-1}^{+}\\
u_{1,-1}^{-}& v_{1,-1}^{-}\\
\end{array}%
\right)^{-1}\left(%
\begin{array}{cc}
\vartheta_{1}^{+} &\vartheta_{1}^{-} \\
\zeta_{1}^{+} &\zeta_{1}^{-} \\
\end{array}%
\right)\\
&& \mathcal{M}_{34}=\left(%
\begin{array}{cc}
 \vartheta_{-1}^{+} &\vartheta_{-1}^{-}\\
  \zeta_{-1}^{+} & \zeta_{-1}^{-} \\
\end{array}%
\right)^{-1}\left(%
\begin{array}{cc}
 u_{-1,1}^{+} &  v_{-1,1}^{+}\\
 u_{-1,1}^{-} & v_{-1,1}^{-} \\
\end{array}%
\right)\\
&& \mathcal{M}_{45}=\left(%
\begin{array}{cc}
 u_{-1,2}^{+} &  v_{-1,2}^{+}\\
 u_{-1,2}^{-} & v_{-1,2}^{-}\\
\end{array}%
\right)^{-1}\left(%
\begin{array}{cc}
  e^{\textbf{\emph{i}}p_{x5} d_{2}} & e^{-\textbf{\emph{i}}p_{x5} d_{2}} \\
  z_{5} e^{\textbf{\emph{i}}p_{x5} d_{2}}  & -z_{5}^{\ast} e^{-\textbf{\emph{i}}p_{x5} d_{2}}  \\
\end{array}%
\right).
\end{eqnarray}
These will enable us to compute the reflection and transmission amplitudes
\begin{equation}\label{eq 63}
 t_{m}=\frac{1}{\tilde{m}_{11}}, \qquad  r_{m}=\frac{\tilde{m}_{21}}{\tilde{m}_{11}}.
\end{equation}
More explicitly, we have for transmission
\beq
t_{m}=
\frac{e^{id_{2}\left(p_{x1}+p_{x5}\right)}\left(1+z_{5}^{2}\right)\left(\vartheta_{1}^{-}\zeta_{1}^{+}+\vartheta_{1}^{+}\zeta_{1}^{-}
\right)}{f_{2}^{+}\left(f_{1}^{-}\mathcal{L}_{1}+if_{2}^{-}\mathcal{L}_{2}\right)+f_{1}^{+}\left(f_{2}^{-}\mathcal{L}_{3}+if_{1}^{-}
\mathcal{L}_{4}\right)}\mathcal{D}
\eeq
where the quantities $\mathcal{D}$, $\mathcal{L}_{1}$, $\mathcal{L}_{2}$, $\mathcal{L}_{3}$ and $\mathcal{L}_{4}$ are defined by
 \begin{eqnarray}
\mathcal{D}&=&
\left(u_{-1,1}^{-}v_{-1,1}^{+}-u_{-1,1}^{+}v_{-1,1}^{-} \right)
\left(u_{1,-2}^{+}v_{1,-2}^{-}-u_{1,-2}^{-}v_{1,-2}^{+}
\right)\\
 \mathcal{L}_{1}&=&
\vartheta_{-1}^{-}\zeta_{1}^{+}\mathcal{F}\mathcal{G}
-\vartheta_{1}^{-}\zeta_{-1}^{+}\mathcal{K}\mathcal{J}\\
\mathcal{L}_{2}&=&\left(\zeta_{1}^{+}\zeta_{-1}^{-}-\zeta_{1}^{-}\zeta_{-1}^{+}\right)\mathcal{F}\mathcal{J}\\
 \mathcal{L}_{3}&=&
\vartheta_{-1}^{+}\zeta_{1}^{-}\mathcal{F}\mathcal{G}-\vartheta_{1}^{+}\zeta_{-1}^{-}\mathcal{K}\mathcal{J}\\
\mathcal{L}_{4}&=&=\left(\vartheta_{1}^{+}\vartheta_{-1}^{-}-\vartheta_{1}^{-}\vartheta_{-1}^{+}\right)\mathcal{K}\mathcal{G}
\end{eqnarray}
and
\begin{eqnarray}
\mathcal{F}&=&\left[u_{1,-1}^{+}v_{1,-2}^{-}-u_{1,-2}^{-}v_{1,-1}^{+}-
z_{1}\left(u_{1,-1}^{+}v_{1,-2}^{+}-u_{1,-2}^{+}v_{1,-1}^{+}\right)\right]\\
\mathcal{G}&=&\left[u_{-1,1}^{-}v_{-1,2}^{+}-u_{-1,2}^{+}v_{-1,1}^{-}
+z_{5}\left(u_{-1,1}^{-}v_{-1,2}^{-}-u_{-1,2}^{-}v_{-1,1}^{-}\right)\right]\\
\mathcal{K}&=&\left[u_{1,-1}^{-}v_{1,-2}^{-}-u_{1,-2}^{-}v_{1,-1}^{-}-z_{1}\left(u_{1,-1}^{-}v_{1,-2}^{+}-u_{1,-2}^{+}v_{1,-1}^{-}\right)\right]\\
\mathcal{J}&=&\left[u_{-1,1}^{+}v_{-1,2}^{+}-u_{-1,2}^{+}v_{-1,1}^{+}+z_{5}\left(u_{-1,1}^{+}v_{-1,2}^{-}-u_{-1,2}^{-}v_{-1,1}^{+}\right)\right]
\end{eqnarray}

Actually what we need are exactly the
transmission $T_m$ and reflection $R_m$ probabilities. These
can be obtained using the electric current density $J$
corresponding to our system. From our previous Hamiltonian,
we can show incident, reflected and transmitted current take the form
\begin{eqnarray}
&& J_{\sf {inc,m}}=  e\upsilon_{F}(\psi_{1}^{+})^{\dagger}\sigma
_{x}\psi_{1}^{+}\\
 && J_{\sf {ref,m}}= e\upsilon_{F} (\psi_{1}^{-})^{\dagger}\sigma _{x}\psi_{1}^{-}\\
 && J_{\sf {tra,m}}= e\upsilon_{F}\psi_{5}^{\dagger}\sigma _{x}\psi_{5}.
\end{eqnarray}
These can be used to write the reflection and transmission probabilities as 
\begin{equation}
  T_{m}= \frac{p_{x5}}{p_{x1}}|t_{m}|^{2}, \qquad
  R_{m}=|r_{m}|^{2}.
\end{equation}

The physical outcome of particle scattering through the double
triangular barrier depends on the energy of the incoming particle.
We numerically evaluate the transmission probability $T_{m}$ as a
function of structural parameters of the graphene double triangular barrier
with a perpendicular magnetic field, including the energy
$\epsilon$, the $y$-component of the wave vector $k_{y}$, the magnetic
field $B$, the energy gap $\mu$ and the applied potentials $v_{1}$ and
$v_{2}$. The results are shown in Figures \ref{figm1}, \ref{figm2}
and \ref{figm3}. In addition to the expected above-barrier full
transmission for some values of $\epsilon l_{B}$ and $v_{2}l_{B}$.\\

 \begin{figure}[h!]
\centering
\includegraphics[width=8cm, height=5cm]{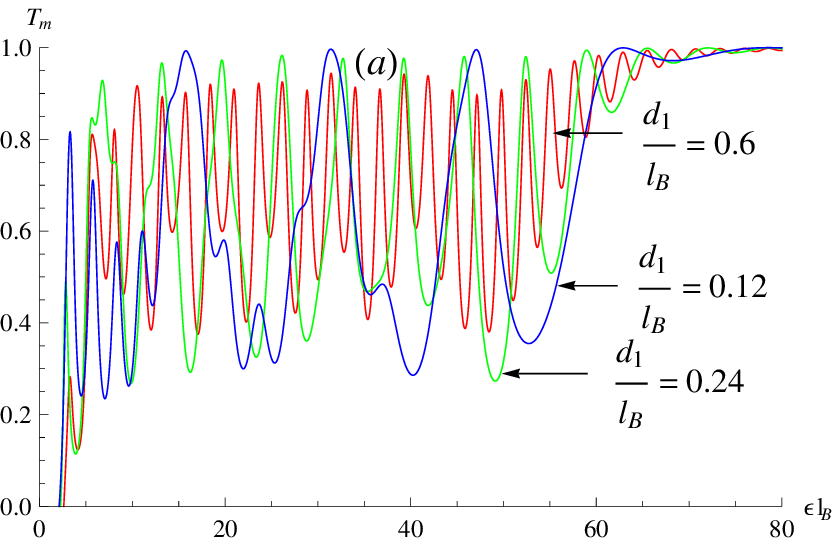}\ \ \ \
\includegraphics[width=8cm, height=5cm]{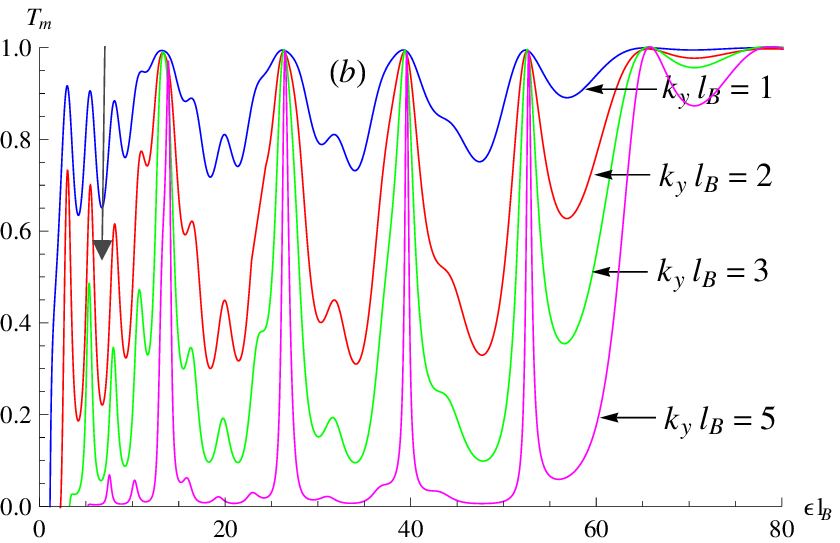}\\
 \caption{\sf{(Color online) Transmission probability $T_{m}$ for the magnetic barrier as a function of energy
 $\epsilon l_{B}$ with $\frac{d_{2}}{l_{B}}=1.5$,
 $v_{1}l_{B}=60$, $v_{2}l_{B}=0$ and $\mu l_{B}=0$. (a) the parameters: $k_{y}l_{B}=2$ and
$\frac{d_{1}}{l_{B}}=\{0.12, 0.24, 0.6\}$. (b) the parameters:
$\frac{d_{1}}{l_{B}}=0.12$ and $k_{y}l_{B}=\{1, 2, 3,
5\}$}}\lb{figm1}
\end{figure}
We note that in Figure \ref{figm1}a), when the energy is less
than the height of the potential barrier $\epsilon
l_{B}<k_{y}l_{B}+\frac{d_{1}}{l_{B}}$, we have zero transmission.
In the second interval
$k_{y}l_{B}+\frac{d_{1}}{l_{B}}\leq\epsilon l_{B}\leq v_{1}l_{B}$
the third zone contains
oscillations. Finally the interval $\epsilon
l_{B}>v_{1}l_{B}$ contains the usual high energy barrier
oscillations and asymptotically goes to unity at high energy.
Figure \ref{figm1}b) shows the transmission spectrum for
different wave vector $k_{y}l_{B}$, the energy gap $\mu l_{B}$ is
zero and $v_{2}l_{B}=0$. We see that if we increase the wave vector $k_{y}l_{B}$
the zone of zero transmission increases following the condition
$\epsilon l_{B}<k_{y}l_{B}+\frac{d_{1}}{l_{B}}$. In the second
interval the transmission oscillates between the value of the total transmission
and zero as $k_{y}l_{B}$ increases. Finally in the interval $\epsilon
l_{B}>v_{1}l_{B}$ the transmission increases. \\

\begin{figure}[h!]
\centering
\includegraphics[width=8cm, height=5cm]{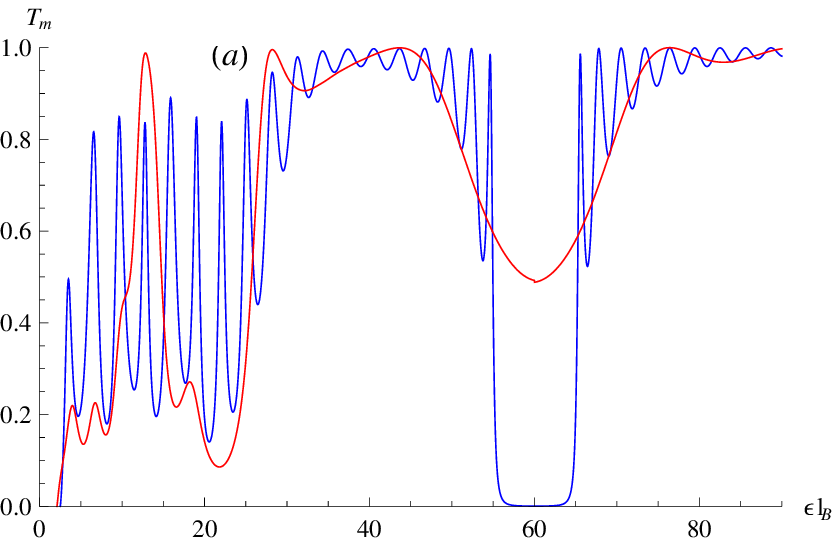}\ \ \ \
\includegraphics[width=8cm, height=5cm]{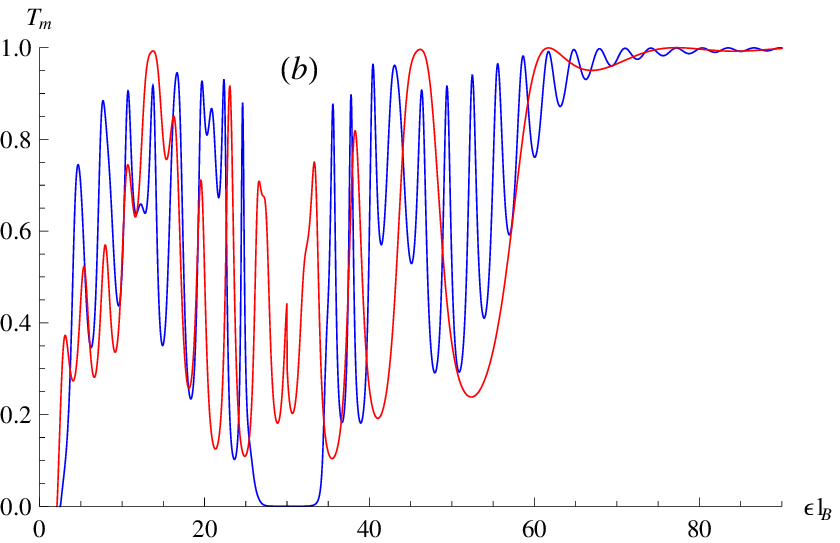}\\
 \caption{\sf{(Color online) Transmission probability $T_{m}$ for the magnetic barrier  as a function of energy
 $E$ with  $\frac{d_{1}}{l_{B}}=0.1$ color red, $\frac{d_{1}}{l_{B}}=0.5$ color blue, $\frac{d_{2}}{l_{B}}=1.5$, $\mu l_{B}=4$ and $k_{y} l_{B}=2$.
 a) the parameters:
 $v_{1} l_{B}=30$ , $v_{2} l_{B}=60$. b) the parameters: $v_{1}l_{B}=60$ , $v_{2}l_{B}=30$
.}}\lb{figm2}
\end{figure}
On the other hand, if we keep the same well region and cancel both the
applied magnetic field and mass term in the well region, the
series of potentials behave like a simple double barrier with the
same effective mass $k_{y}$. Thus, in this case we
reproduce exactly the transmission obtained in \cite{Alhaidari},
for the massive Dirac equation with $m = k_{y}$. Let us treat the triangular
double barrier case when $v_{2}<v_{1}$ and
$v_{2}>v_{1}$. In both cases, the transmission is plotted in
Figure \ref{figm2}: In Figure \ref{figm2}a) $v_{2}>v_{1}$ we
distinguish five different zones characterizing the behavior of
the transmission coefficient :
\begin{itemize}
 \item 
The first is determined by the greater effective mass,
namely $\epsilon l_{B}<k_{y}l_{B}+\frac{d_{1}}{l_{B}}$.
\item  The second identifies with the lower Klein energy zone characterized
by resonances and $k_{y}l_{B}+\frac{d_{1}}{l_{B}}<\epsilon
l_{B}<v_{1} l_{B}$. Here we have full transmission at some
specific energies despite the fact that the particle energy is
less than the height of the barrier. As $d_{1}/l_{B}$ increases,
the oscillations in the Klein zone get reduced. This strong
reduction in the transmission in the Klein zone seem to suggest
the potential suppression of the Klein tunneling as we increase
$d_{1}/l_{B}$.
\item 
 The third zone $v_{1}l_{B}<\epsilon
l_{B}<v_{2}l_{B}-k_{y}l_{B}-\frac{\mu l_{B}}{2}$ is a window where
the transmission oscillates around the value of the total
transmission.
\item  The fourth zone defined by $v_{2}l_{B}-k_{y}l_{B}-\frac{\mu
l_{B}}{2}<\epsilon l_{B}<v_{2}l_{B}+k_{y}l_{B}+\frac{\mu
l_{B}}{2}$ is a window where the transmission is almost zero.
\item
 The fifth zone $\epsilon
l_{B}>v_{2}l_{B}+k_{y}l_{B}+\frac{\mu l_{B}}{2}$ contains
oscillations, the transmission converges towards unity.
\end{itemize}
Contrary to the case $v_{1}>v_{2}$, see Figure \ref{figm2}b) we distinguish fourth different zones
characterizing the behavior of the transmission coefficient:
\begin{itemize}
 \item 
Compared to Figure \ref{figm2}a), the behavior in the
first zone is the same as in in Figure
\ref{figm2}a). 
\item Concerning the zones $k_{y}l_{B}-\frac{d1
}{l_{B}}<\epsilon l_{B}<v_{2}l_{B}-k_{y}l_{B}-\frac{\mu l_{B}}{2}$
and $v_{2}l_{B}+k_{y}l_{B}+\frac{\mu l_{B}}{2}<\epsilon
l_{B}<v_{1}l_{B}$ the transmission oscillates similarly to Figure
\ref{figm2}a). 
\item In the zone $v_{2}l_{B}-k_{y}l_{B}-\frac{\mu
l_{B}}{2}<\epsilon l_{B}<v_{2}l_{B}+k_{y}l_{B}+\frac{\mu
l_{B}}{2}$, one can see that both curves start from zero
transmission and oscillate while the valley gets wider as
$d_{1}/l_{B}$ decreases. 
\item Finally zone $\epsilon l_{B}>v_{1}l_{B}$
the transmission oscillate to reach the total transmission.
\end{itemize}

\begin{figure}[h!]
\centering
\includegraphics[width=8cm, height=5cm]{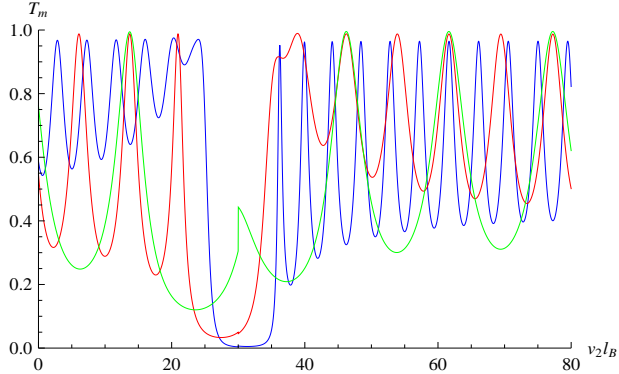}\\
 \caption{\sf{(Color online) Transmission probability $T_{m}$ for the magnetic barrier  as a function of
 potential
 $v_{2}l_{B}$ with   $\frac{d_{1}}{l_{B}}=0.1$ color green, $\frac{d_{1}}{l_{B}}=0.2$ color red, $\frac{d_{1}}{l_{B}}=0.34$ color blue,
 $\frac{d_{2}}{l_{B}}=1.5$, $\mu l_{B}=4$, $k_{y} l_{B}=2$,
 $v_{1}l_{B}=60$ and $\epsilon l_{B}=30$.
.}}\lb{figm3}
\end{figure}

 It is worth to analyze the transmission versus
the potential $v_{2}l_{B}$. In doing so, we choose a fixed value of
$d_{1}/l_{B}$ to present Figure \ref{figm3}. It is clear that two transmission curves increase
while $d_{1}/l_{B}$ decreases in the intermediate zone.

\section{Conclusion}

We have considered a model to describe over-barrier
electron emission from the edge of monolayer graphene through a
triangular electrostatic double barriers in addition to a magnetic field
in graphene. 
To underline the behavior of our system, we have separately considered two parts: first including
static barrier and second deal with magnetic barrier. In both cases, we have set the materials
needed to analytically determine and numerically analyze the transmission probability. These have
been done by solving
 the eigenvalue equation 
to end up with the solutions of the energy spectrum in terms of different
physical parameters involved in the Hamiltonian system.

By using the continuity of the wavefunctions at the interfaces between different regions
inside and outside the barriers we have ensured conservation of the local current density and
derived the relevant transport coefficients of the present system. Specifically, using the
transfer matrix method, we have analyzed the corresponding
transmission coefficient and determined how the transmission
probability is affected by various physical parameters. In particular for static barrier,
the resonances were seen in different regions as
well as the Klein tunneling effect. 

Subsequently, we have analyzed the same system but this time by taking  
into account the presence of an inhomogeneous magnetic field. Using boundary conditions, we have split
the energy into three domains and then calculated the transmission probability in each case. In each
situation, we have discussed the transmission at resonances that characterize each region and stressed the
importance of our results. 

\section*{Acknowledgments}

The generous support provided by the Saudi Center for Theoretical Physics (SCTP)
is highly appreciated by all authors. AJ acknowledges partial support by King Faisal University
while HB acknowledge the support of King Fahd University of Petroleum and minerals under
research group project RGxxxx.



\begin{thebibliography}{99}

\bibitem{Novoselov} K. S. Novoselov, A. K. Geim, S. V. Morozov, D. Jiang, Y. Zhang, S. V.
Dubonos, I. V. Grigorieva and A. A. Firsov, {Science} {306}, 666
(2004).

\bibitem{Stander} N. Stander, B. Huard and D. Goldhaber-Gordon, Phys. Rev. Lett. 102, 026807 (2009).

\bibitem{Katsnelsonn} M. I. Katsnelson, K. S. Novoselov and A. K. Geim, Nature Phys. 2, 620 (2006).

\bibitem{Sevinçli} H. Sevincli, M. Topsakal and S. Ciraci, Phys. Rev. B 78,   245402 (2008).

\bibitem{Sevin} H. Sevinžcli, M. Topsakal and S. Ciraci,
Phys. Rev. B 78, 245402 (2008).
 \bibitem{Masir} M. R. Masir, P.
Vasilopoulos and F. M. Peeters, New J. Phys. 11, 095009 (2009).
\bibitem{DellAnna} L. DellAnna and A. De Martino, Phys. Rev. B 79, 045420
(2009).
\bibitem{Mukhopadhyay} S. Mukhopadhyay, R. Biswas and C. Sinha, Phys.
Status Solidi B 247, 342 (2010).
\bibitem{Choubabi} E. B. Choubabi, M. El
Bouziani and A. Jellal, Int. J. Geom. Meth. Mod. Phys. 7, 909
(2010).
\bibitem{Mekkaoui} H. Bahlouli, E. B. Choubabi, A. Jellal and M.
Mekkaoui, J. Low Temp. Phys. 169, 51 (2012).

\bibitem{Jellal} A. Jellal and A. El Mouhafid, J. Phys. A:
Math. Theo. 44, 015302 (2011).
\bibitem{Tworzydlo} J. Tworzydlo, B.
Trauzettel, M. Titov, A. Rycerz and C. W. J. Beenakker, Phys. Rev.
Lett. 96, 246802 (2006).
\bibitem{Alhaidari} A. D. Alhaidari, H. Bahlouli
and A. Jellal, 
Advances in Mathematical Physics 2012,  ID 762908 (2012).


\bibitem{HBahlouli} H. Bahlouli, E.B. Choubabi, A. El Mouhafid and A. Jellal, Solid
State Communications 151, 1309 (2011).

\bibitem{Matulis} A. Matulis, F. M. Peeters, P. Vasilopoulos, Phys. Rev. Lett.
72, 1518 (1994).

\bibitem{Ramezani} M. Ramezani Masir, P. Vasilopoulos, F. M. Peeters, Phys.
Rev. B 82, 115417 (2010).

\bibitem{Tworzydlo} J. Tworzydlo, B. Trauzettel, M. Titov, A. Rycerz and C. W. J. Beenakker, Phys. Rev. Lett. 96,
246802 (2006).

\bibitem{Berry} M. V. Berry and R. J. Modragon, Proc. R. Soc. London Ser. A 412, 53 (1987).



\bibitem{Abramowitz}  M. Abramowitz and I. Stegum, Handbook of Integrabls, Se ries and
Products, (Dover, New York, 1956).
\bibitem{Gonzalez} L. Gonzalez-Diaz and V. M. Villalba, Phys. Lett. A 352, 202
(2006).






















\end{thebibliography}
\end{document}